# Exploring one aspect of pedagogical content knowledge of teaching assistants using the Conceptual Survey of Electricity and Magnetism

Nafis I. Karim,[1] Alexandru Maries,[2] and Chandralekha Singh[1]

[1]*Department of Physics and Astronomy, University of Pittsburgh, Pittsburgh, Pennsylvania 15260, USA*
[2]*Department of Physics, University of Cincinnati, Cincinnati, Ohio 45221, USA*

The Conceptual Survey of Electricity and Magnetism (CSEM) has been used to assess student understanding of introductory concepts of electricity and magnetism because many of the items on the CSEM have strong distractor choices which correspond to students' alternate conceptions. Instruction is unlikely to be effective if instructors do not know the common alternate conceptions of introductory physics students and explicitly take into account common student difficulties in their instructional design. Here, we discuss research involving the CSEM to evaluate one aspect of the pedagogical content knowledge of teaching assistants (TAs): knowledge of introductory students' alternate conceptions in electricity and magnetism as revealed by the CSEM. For each item on the CSEM, the TAs were asked to identify the most common incorrect answer choice selected by introductory physics students if they did not know the correct answer after traditional instruction. Then, we used introductory student CSEM post-test data to assess the extent to which TAs were able to identify the most common alternate conception of introductory students in each question on the CSEM. We find that the TAs were thoughtful when attempting to identify common student difficulties and they enjoyed learning about student difficulties this way. However, they struggled to identify many common difficulties of introductory students that persist after traditional instruction. We discuss specific alternate conceptions that persist after traditional instruction, the extent to which TAs were able to identify them, and results from think-aloud interviews with TAs which provided valuable information regarding why the TAs sometimes selected certain alternate conceptions as the most common but were instead very rare among introductory students. We also discuss how tasks such as the one used in this study can be used in professional development programs to engender productive discussions about the importance of being knowledgeable about student alternate conceptions in order to help students learn. Interviews with TAs engaged in this task as well as our experience with such tasks in our professional development programs suggest that they are beneficial.

DOI: 10.1103/PhysRevPhysEducRes.14.010117

## I. INTRODUCTION

### A. Graduate teaching assistants

Graduate students in physics across the United States have been playing an important role in educating the next generation of students for a long time. In particular, in the U.S. it is quite common for physics graduate teaching assistants (TAs) to teach introductory physics recitation or lab sections which typically have lower enrollments than the "lecture" component of the course (20–40 compared to 100 or more in a lecture). In addition to the graduate TAs, in the last two decades, undergraduate TAs [sometimes referred to as learning assistants (LAs)] have also played a role in educating students by, e.g., assisting faculty members in teaching large classes. Appropriate professional development of these TAs to help them perform their duties effectively is an important task. Physics education researchers have been involved in research on identifying common beliefs and practices among physics TAs that have implications for effective teaching [1–9]. For example, research suggests that sometimes graduate TAs struggle to understand the value of thinking about the difficulty of a problem from an introductory student's perspective and think that if they know the material and can explain it to their students in a clear manner, it will be sufficient to help their students learn [1,3] [note that throughout this paper, "student(s)" refers to introductory physics student(s), and "recitations or labs" refers to "introductory physics recitations or labs"]. Also, while graduate TAs are able to recognize useful solution features and articulate why they are important when looking at sample student solutions provided to them, they do not necessarily include those features in their own written solutions [4–6]. Moreover, the TAs do not always engage in grading practices that are conducive to helping students learn expertlike problem solving strategies and develop a coherent understanding of physics [7,8].

It is also important to keep in mind that TAs may be given varying amounts of freedom regarding how to perform their teaching duties, depending on the instructor. However, discussions with the TAs who participated in this study and others from the University of Pittsburgh (Pitt) suggest that except for broader guidelines such as whether to discuss homework problems followed by a quiz or whether to have group problem solving [10–16] followed by a quiz in the recitation, the TAs often have considerable flexibility in how to perform their recitation duties. For example, many instructors meet with the TA only briefly at the beginning of the semester to outline general guidelines, e.g., answer student questions on the homework, solve problems on the board, and the TAs are left to their own devices for the rest of the semester, except for some communication with the course instructor via email or during the grading of the exams. Thus, if TAs are knowledgeable about effective instructional approaches, they can make a significant contribution to students' learning of physics in the recitations because they often have sufficient flexibility to lead the recitation in a manner that they think is conducive to student learning.

To help TAs learn about effective pedagogy, many institutions offer professional development programs which are sometimes discipline specific [9,17–19]. For more information about professional development programs and research on recruiting and educating future teachers, see Ref. [9] and references therein. The effectiveness of these professional development programs can be enhanced if those leading them are knowledgeable about TAs' conceptions regarding students' difficulties [20]. For example, TAs may be largely unaware of certain student alternate conceptions. If professional development instructors preparing TAs discuss students' alternate conceptions and engage the TAs in discussions about how to help them learn, the TAs may be better prepared to conduct their teaching duties. It is even possible that in order to convince the TAs, the professional development instructors may need to have TAs reflect upon quantitative data on student performance, which show that those alternate conceptions are common. This type of activity in TA professional development programs has the potential to enhance TAs' teaching effectiveness as they design, adopt, and adapt activities to help students develop a robust knowledge structure. Similarly, if TA professional development instructors are aware that TAs know about certain student alternate conceptions, those can only be discussed briefly.

Thus, by focusing on what TAs know and do not know and gradually building and strengthening different aspects of their *pedagogical content knowledge* (PCK) [21,22] (more about PCK in the next section), they can be guided to learn and implement effective pedagogies. These considerations motivated us to carry out the research study discussed here using the Conceptual Survey on Electricity and Magnetism (CSEM), which is one of the many assessment tools often used to evaluate students' conceptual understanding of introductory concepts [23]. The goal of the present study was to evaluate TAs' knowledge of student alternate conceptions in electricity and magnetism as revealed by the CSEM. For each item on the CSEM, the TAs were asked to identify the most common incorrect answer choice (MCI) selected by students after traditional instruction. This exercise was followed by a class discussion with the TAs related to this task, including the importance of knowing student difficulties and addressing them effectively in order for learning to be meaningful. We have found that this type of activity in a TA professional development course engenders a rich discussion about student difficulties and promotes the importance of thinking about their difficulties from students' perspective in order to bridge the gap between teaching and learning. More information about potential uses of this type of activity in TA professional development is provided in Section IV.

### B. Pedagogical content knowledge

There are several theoretical frameworks that inspire our research. These theoretical frameworks emphasize the importance of instructors familiarizing themselves with students' common difficulties in order to scaffold their learning with appropriately designed curricula and pedagogies. In the context of this study, they point to the importance of being knowledgeable about student difficulties in order to help students learn better. For example, Piaget [24] emphasized "optimal mismatch" between student ideas and instructional design for cognitive conflict and desired assimilation and accommodation of knowledge, and others have put forth similar ideas [25].

Being knowledgeable about student alternative conceptions related to a particular topic and using them as resources in instructional design is one aspect of what Shulman defined as pedagogical content knowledge (PCK) [21,22]. Shulman defines PCK as the subject matter knowledge *for teaching*. In other words, PCK is a form of practical knowledge used by experts to guide their pedagogical practices in highly contextualized settings. Shulman writes "Within the category of pedagogical content knowledge, I include […] the most useful forms of representation of those ideas, the most powerful analogies, illustrations, examples, explanations, and demonstrations—in a word, the ways of representing and formulating the subject that make it comprehensible to others." In addition, according to Shulman, PCK also includes "an understanding of what makes the learning of specific topics difficult," or, in other words, knowledge of the common difficulties that students have in learning a specific topics [21]. Shulman developed the concept of PCK in response to the growing trend of proliferating general educational research in teacher preparation programs. The development of PCK was in part due to Shulman's previous research on the reasoning processes of physicians [26], which he found to be domain specific and contrary to the general assumption that certain physicians possess a general trait of diagnostic acumen which makes them better diagnosticians than others. Shulman generalized this observation to conclude that good teachers not only possess domain specific knowledge, but also possess more practical knowledge about teaching that is domain specific (i.e., PCK). Shulman therefore encouraged research on teachers' PCK and the types of teacher preparation programs that are likely to improve and/or develop teachers' PCK. Since Shulman introduced the concept of PCK, much has been written about it [27–42]. For example, Grossman [29] includes PCK as one of the "four general areas of teacher knowledge [which are] the cornerstones of the emerging work on professional knowledge for teaching: general pedagogical knowledge, subject matter knowledge, pedagogical content knowledge, and knowledge of context" and argues that PCK (as opposed to their subject matter knowledge) generally has the greatest impact on teachers' classroom activities. Others have also stressed the importance of PCK in shaping instructional practice and discuss professional development programs which take PCK into account [36,37]. For example, Borko and Putman [37] describe the Cognitively Guided Instruction Project, a multiyear program of curriculum development, professional development, and research that has shown "powerful evidence that experienced teachers' pedagogical content knowledge and pedagogical content beliefs can be affected by professional development programmes."

Given the importance of PCK in shaping instructional practices, it is not surprising that researchers have attempted to document teachers' PCK [31,33,34] and others have attempted to document the development of teachers' PCK [35,38]. However, these tasks are challenging to carry out for multiple reasons such as the fact that much of the knowledge teachers have of their practice is tacit [39,40], or the fact that although there is a general consensus among researchers on PCK as a construct, its boundaries are not clearly delineated [41]. Also, extended observations are needed in order to recognize when teachers' PCK is instantiated in their practice [31]. To overcome some of these challenges, researchers have often used multimethod approaches to investigate teachers' PCK. For example, observational data are not sufficient because a teacher may use only a small portion of the representations they have at their disposal. In addition, observations do not provide insight into teachers' instructional decisions—we see what they are doing, but do not know why. Partly due to these issues, Loughran et al. [31] used both classroom observations and follow-up interviewing of teachers. The interviews encouraged teachers to articulate their knowledge and explored alternative representations that the teachers did not use during the teaching sessions. This investigative approach is quite time consuming to both carry out and analyze since both the observations and interviews provide lengthy qualitative data which require coding and analysis. Baxter and Lederman [42] provide a review of methods and techniques for studying PCK and the subject matter knowledge of teachers.

Partly due to all of the difficulties in carrying out an involved investigation of PCK, we developed a relatively straightforward method for delving into one particular aspect of PCK, namely, knowledge of student difficulties with particular topics. This method makes use of standardized conceptual multiple-choice tests developed by physics education researchers and quantitative data from students taking these tests. Teachers are provided with a copy of a particular test (e.g., CSEM), and for each item on the test they are asked to select what they expect would be the MCI selected by students after traditional instruction in a relevant topic. Then, quantitative student data are used to quantify the extent to which teachers are knowledgeable about common student difficulties revealed by students' MCIs after traditional instruction. Previous research with K-12 teachers [20] found that on items which have a strong distractor (e.g., MCI), there is a large difference in learning gains between students taught by teachers who could identify the alternate conceptions and students taught by teachers who could not. Therefore, it is valuable to explore the extent to which teachers are knowledgeable about student alternate conceptions on items drawn from well-designed standardized tests.

Two prior research studies conducted with TAs using the method described in the preceding paragraph used the Force Concept Inventory (FCI) [43,44] and Test of Understanding Graphs in Kinematics (TUG-K) [45,46]. The main findings from these studies are as follows:

- TAs were able to identify student MCIs in certain contexts, but struggled to identify them in other contexts. For example, for the FCI, 84% of the TAs identified students' alternate conception related to Newton's 3rd law in the typical context (car colliding with truck), but only 40% of the TAs identified it in a less typical context (car pushing truck and speeding up).
- TAs sometimes expected certain answer choices to be the MCIs, when instead those answer choices were selected by very few students.
- Think-aloud interviews with TAs engaged in the task of determining students' MCIs suggested that the TAs were reflective and often had reasonable thoughts regarding how students may be reasoning about the questions. Interviews also suggested that the TAs were sometimes distracted by certain answer choices that were not common among students and reasoned that those answer choices would be the MCIs.

In this study, we extend our previous work and use the CSEM to investigate the extent to which TAs enrolled in a semester-long professional development course are knowledgeable about the MCIs of students related to electricity and magnetism after traditional instruction as revealed by the CSEM. We also discuss how being aware of TAs' knowledge of MCIs can be useful in designing effective professional development programs.

## II. METHODOLOGY

### A. Participants

The participants in this study were 81 first year graduate students (three separate cohorts) enrolled in a semester long mandatory pedagogy oriented TA professional development course at Pitt, which meets once a week for two hours. The graduate student population at Pitt is consistent with that of a typical research-based state university. The TAs teach recitations and labs, typically in a traditional manner. In the recitations, the TAs primarily answer student questions, solve problems on the board and give students a quiz in the last 10–20 min. In the labs, the TAs start by demonstrating the procedures needed for that lab and the students closely follow the detailed procedures written in the lab manual.

Since this is the first and last pedagogy-oriented semester long course most physics graduate TAs at Pitt will ever take, it is designed to help them become more effective teachers in general. During the course, they get a general overview of cognitive research and PER during 1 two-hour session and discuss with each other and reflect upon their instructional implications. The TAs are also introduced to curricula and pedagogies based on PER which emphasize the importance of being knowledgeable about students' difficulties in order to help them develop expertise in physics. Each week, TAs complete various reflective exercises designed to help them perform their TA duties in a student-centered manner. For example, in one class, they discuss how to write effective problem solutions for physics classes and what features should be included in solutions they hand out to students and why [4–6]. In another class, they are given sample student solutions and asked to grade them individually and in groups, followed by a discussion about how to grade students using a rubric to help them learn better [7,8]. In the second half of the semester, each TA also leads an interactive discussion of the solution of a physics problem in the class in the manner in which they would lead a discussion if teaching students and receive feedback from the other TAs in the course (who are asked to pretend to be students and ask questions) and the instructor. Thus, the TA professional development course (which is required of all first-year graduate students) is not focused on helping the TAs implement physics education research (PER) based curricula in specific recitations or labs (e.g., University of Washington tutorials [47]), but is a general introduction to pedagogical issues in physics teaching and learning.

This study focuses on issues related to the professional development of TAs who teach recitations and labs and typically have a closer association with students than the course instructors and, thus, they may be in an even better position (compared to the course instructors) to help students learn if the TAs are versed in effective pedagogy. At Pitt, the TAs generally hold regular office hours and interact with students in the physics resource room where they help students with any questions related to their physics courses. In addition, recitation class sizes are usually much smaller than the sizes of lecture classes taught by instructors. Therefore, TAs who are knowledgeable about student difficulties related to electricity and magnetism concepts can play a significant role in improving student understanding of these concepts using appropriate curricula or pedagogies and they can address students' difficulties directly in their interactions with students.

In addition to the quantitative study, we conducted think-aloud interviews [48] with 11 TAs. Because of the availability of the TAs for individual interviews, four of the interviewed TAs had participated in the quantitative study (they were in the TA professional development course in which the quantitative study was carried out) but seven were not. We also note that for the TAs who participated in the quantitative study, at least one year had passed before they were interviewed. Thus, the questions in the CSEM PCK task carried out in the TA professional development course were not fresh in their mind at the time of the interviews. Each of the 11 TAs had at least one semester of teaching experience in recitations. We did not find any qualitative differences in the reasoning of the TAs whether they had participated in the quantitative study earlier or not. More details about the interviews are provided in Sec. II C.

### B. Materials

The materials used in this study are the CSEM, which was given to the TAs in the TA professional development course as explained below, the post-instruction algebra-based students' data (printed in Ref. [23]) that were collected over a period of four years from an average of 388 students, the quantitative data and the interview data obtained from the TAs. These data were used to determine students' MCIs on each item on the CSEM, to assess TA knowledge of student alternate conceptions, and to understand the reasoning TAs use when selecting certain incorrect answers as the MCIs.

### C. Methods

In the quantitative study, the TAs were provided with the CSEM and, for each item on the CSEM, they were asked to identify what they expected to be the MCI of students if they did not know the correct answer in a post-test (after traditional instruction in relevant concepts). We refer to this task as the CSEM-related PCK task. We note that the task given to TAs was framed such that they had to identify the MCI for each multiple-choice question that students would select *after* traditional instruction if they did not know the correct answer (rather than *before* instruction), because it was considered that it is more important for TAs to be aware of alternate conceptions that persist after traditional instruction. Also, our previous research suggests that both TAs and instructors [1,44] are often reluctant to contemplate students' conceptions before instruction. We note that it does not make a significant difference whether the question is phrased about students' difficulties with each question in the post-test or pretest because the MCIs of students rarely changed after traditional instruction [23]. Also, an analysis of the pre- and post-test data in Ref. [23] for each item on the CSEM suggests that the percentage of students who had a certain alternate conception either decreased after instruction or remained roughly the same. Since we asked the TAs to identify the alternate conceptions after traditional instruction, we performed data analysis using the post-test data in Ref. [23].

In years two and three of the study, the researchers also asked TAs to predict the percentage of students who would answer each question on the CSEM correctly in a post-test after traditional instruction in relevant concepts. We investigated TA data from each year separately and found very few differences between the different years. Therefore, all the data were combined (for TAs' predictions on the percentage of students answering each question correctly, only years two and three were combined because this question was not asked during the first year). Each year, after the TAs completed the CSEM-related PCK task, there was a full class discussion about the tasks and why knowledge of common student difficulties is critical for teaching and learning to be effective in general. The TAs were not prompted to explain their reasoning for their choices in written responses, but in the class discussion, certain items on the CSEM were discussed in detail and TAs discussed their reasoning for why they expected certain incorrect answer choices to be MCIs of students.

In order to obtain an in-depth account of TAs' reasoning (related to why they expected certain answer choices to be the MCIs), think-aloud interviews were conducted with 11 TAs. Certain questions were selected from the CSEM the TAs were asked to think aloud and (i) identify the correct answer, and (ii) determine the MCI of introductory students for each question. They were not disturbed during this time unless they became quiet for a long time in which case they were asked to keep talking. After discussing all of the questions selected by the interviewer, if time permitted, the TA was sometimes asked to look back at some of the questions and provide more details about why they expected a particular incorrect answer choice to be the MCI if their reasoning was not clear enough when thinking aloud without being disturbed. The main goal of the interviews was to identify possible reasons why TAs expected that certain answer choices would be the MCIs when in fact those answer choices were not common. Thus, the quantitative data collected were used to identify questions in which this may be occurring and the interviews focused on those questions. For example, on Q2 on the CSEM, roughly half the TAs expected that choice D would be the MCI, but this answer choice was only selected by 11% of students (see Table I).

In order to obtain a quantitative measure of TAs' performance at identifying the alternate conceptions of students, scores were assigned to each TA. A TA who selected a particular incorrect answer choice as the MCI in a particular question received a PCK score which was equal to the fraction of students who selected that particular incorrect answer choice. If a TA selected the correct answer choice as the MCI (a rare occurrence), their data were removed *only* for that specific question because they were explicitly asked to indicate the *incorrect* answer choice which is most commonly selected by students if they did not know the correct answer after traditional instruction in relevant concepts. For example, on question 1, the percentages of algebra-based students who selected A, B, C, D, and E are 4%, 63%, 23%, 7%, and 3%, respectively (as shown in Table I). Answer choice B is correct, thus, the PCK score assigned to TAs for each answer choice if they selected it as the MCI would be 0.04, 0, 0.23, 0.07, and 0.03 (A, B, C, D, and E). The total PCK score a TA would obtain on the task for the entire CSEM can be obtained by summing over all of the questions (this is referred to as "CSEM-related PCK score"). More details on how this was done are provided in the Supplemental Material [49].

We note that the approach used to determine the CSEM-related PCK score weighs the responses of TAs by the fraction of students who selected a particular incorrect response. This weighting scheme was chosen because the more prevalent a student difficulty is, the more important it is for a TA to be aware of it and take it into account in their instruction. Furthermore, this approach also provides a reasonable PCK score when there is more than one common alternate conception. For example, if a question has two incorrect answer choices that are commonly selected by students, e.g., Q29 in which 26% of students selected A and 23% selected B (both incorrect). If all the TAs selected choice A as the MCI, their PCK score would be 100%, but if half the TAs selected A and half select B, their PCK score would be 92.5%.

It is important to clarify that PCK score is *only one* metric of TAs' performance at identifying students' MCIs. In order to get a clear picture of TAs' performance, one needs to look at the percentages of TAs who selected each incorrect answer choice as well as how common those answer choices were.

TABLE I. Questions on the CSEM, percentages of algebra-based students who answered the questions correctly in a post-test, percentages of students who selected each incorrect answer choice ranked from most to least common, the percentage of TAs who selected each incorrect answer choice as the MCI, and average PCK score. To help make the table easier to interpret, answer choices selected by 20% or more students are written in red font. The same answer choices are also written in red font for the TAs. Note that the data were taken from Ref. [23] and the number of students who answered each question varies from 158 to 444. With the exception of four questions, all questions were answered by more than 350 students in the post-test.

| CSEM Item # | Correct Answer | Intro Student Choices | | | | TA Choices | | | | Average PCK Score |
|---|---|---|---|---|---|---|---|---|---|---|
| | | 1st | 2nd | 3rd | 4th | 1st | 2nd | 3rd | 4th | |
| 1 | 63% B | 23% C | 7% D | 4% A | 3% E | 54% C | 19% D | 15% E | 11% A | 14.8 |
| 2 | 42% A | 21% B | 19% E | 11% D | 5% C | 49% D | 25% B | 16% E | 10% C | 14.2 |
| 3 | 76% B | 9% C | 8% D | 5% A | 0% E | 54% A | 26% D | 19% C | 1% E | 6.5 |
| 4 | 40% B | 32% C | 21% D | 5% A | 2% E | 57% C | 22% A | 21% D | 0% E | 23.7 |
| 5 | 32% C | 22% D | 20% B | 14% A | 11% E | 50% D | 32% A | 16% B | 1% E | 19.0 |
| 6 | 67% E | 13% B | 10% C | 7% A | 4% D | 51% A | 34% B | 12% C | 3% D | 9.3 |
| 7 | 31% B | 42% C | 19% A | 5% D | 2% E | 45% A | 43% C | 6% D | 6% E | 26.9 |
| 8 | 53% B | 21% D | 10% C | 8% E | 5% A | 43% D | 27% E | 16% C | 14% A | 13.5 |
| 9 | 52% B | 16% D | 12% C | 10% A | 5% E | 36% D | 23% E | 22% C | 19% A | 11.4 |
| 10 | 35% C | 25% E | 20% B | 12% D | 6% A | 47% B | 25% A | 20% D | 9% E | 15.4 |
| 11 | 33% E | 30% A | 14% B | 13% C | 9% D | 45% A | 26% C | 22% D | 6% B | 19.9 |
| 12 | 67% D | 13% C | 9% A | 8% B | 2% E | 31% C | 29% B | 23% A | 17% E | 8.8 |
| 13 | 51% E | 27% A | 20% B | 1% C | 0% D | 56% A | 42% B | 1% C | 1% D | 23.5 |
| 14 | 16% D | 54% A | 13% E | 9% B | 4% C | 46% A | 24% E | 18% B | 11% C | 30.1 |
| 15 | 24% A | 34% C | 24% B | 9% D | 8% E | 74% C | 17% B | 5% E | 4% D | 29.9 |
| 16 | 32% E | 22% B | 17% D | 13% A | 13% C | 35% A | 25% D | 24% C | 16% B | 15.5 |
| 17 | 51% E | 23% C | 16% B | 6% D | 2% A | 63% C | 18% B | 17% D | 2% A | 18.3 |
| 18 | 47% D | 28% E | 17% C | 4% B | 2% A | 52% E | 37% C | 7% B | 3% A | 21.3 |
| 19 | 34% A | 25% B | 14% C | 11% D | 10% E | 61% B | 18% E | 11% D | 10% C | 19.6 |
| 20 | 17% D | 32% C | 20% B | 18% A | 8% E | 44% C | 25% A | 20% B | 11% E | 23.4 |
| 21 | 44% E | 21% C | 15% A | 8% B | 8% D | 41% C | 31% B | 17% A | 12% D | 14.5 |
| 22 | 32% D | 28% C | 22% A | 11% B | 4% E | 59% C | 22% A | 12% E | 7% B | 22.6 |
| 23 | 45% A | 15% B | 13% C | 11% E | 9% D | 48% D | 20% E | 17% B | 16% C | 11.0 |
| 24 | 25% C | 45% B | 19% D | 8% E | 2% A | 48% D | 45% B | 4% A | 3% E | 29.7 |
| 25 | 48% D | 20% C | 12% B | 11% A | 5% E | 35% E | 25% C | 25% A | 16% B | 11.3 |
| 26 | 49% A | 21% D | 11% B | 6% C | 6% E | 47% C | 29% D | 17% B | 7% E | 11.2 |
| 27 | 40% E | 23% D | 19% A | 8% C | 5% B | 46% A | 24% C | 18% B | 12% D | 14.3 |
| 28 | 40% C | 35% E | 12% B | 8% A | 3% D | 55% E | 32% A | 11% B | 2% D | 23.2 |
| 29 | 23% C | 26% A | 23% B | 19% D | 6% E | 51% B | 26% A | 16% D | 7% E | 22.0 |
| 30 | 48% A | 15% C | 14% D | 9% E | 7% B | 66% D | 17% C | 10% E | 7% B | 13.2 |
| 31 | 26% E | 25% C | 18% A | 17% D | 15% B | 47% D | 20% C | 17% B | 16% A | 18.4 |
| 32 | 18% D | 40% B | 23% A | 16% C | 1% E | 42% A | 31% E | 16% C | 11% B | 17.0 |

Furthermore, interviews with TAs engaged in the PCK task can also shed some light on the reasoning the TAs used, and in this study we have done both.

### D. Research goal: How well do TAs predict students' responses to the CSEM after instruction?

In order to investigate this research goal, we analyzed data pertaining to the following:

- Alternate conceptions which many TAs expected to be common, which were instead rare among students.
- Alternate conceptions which were common among students but were not identified very well by TAs.
- Alternate conceptions that were common among students, which the majority of TAs were able to identify.
- Qualitative results from detailed think-aloud interviews with 11 TAs that focused on what common

reasoning TAs used to select certain answer choices as the MCIs (e.g., answer choices which were not common among students).
- The extent to which TAs were able to predict the difficulty of the questions.

We note the following about the interviews: in general, during the interviews, the TAs were reflective and sometimes thought back to when they were teaching recitations themselves. In some of the questions, they were able to identify the MCI and had good ideas about the common difficulties of students. However, an important goal of the interviews was to identify the reasoning the TAs commonly use when they select answer choices that were not very common among students. Therefore, the discussion focusing on this aspect in a particular question should not be taken as an indication that the interviewed TAs did a poor job at identifying common alternate conceptions of students on those questions.

## III. RESULTS

We note that the MCIs of students are similar for both algebra-based and calculus-based classes (see Ref. [23]). Therefore, the researchers performed the analysis of the CSEM-related PCK performance with the student data from algebra-based classes in Ref. [23] as discussed below.

### A. Performance of TAs in identifying students' alternate conceptions related to the CSEM

Many questions on the CSEM reveal a common student alternate conception [23]. Analysis of the CSEM-related PCK score of the TAs was conducted on each of these questions and the results are displayed in Table I. Table I shows all CSEM items, the percentages of students who answered each question correctly, the percentages of students who selected each incorrect answer choice ranked from most to least common, the incorrect answer choices most commonly selected by TAs (as the MCIs), and the percentages of TAs who selected these answer choices. Correct answers are indicated by the green shading in Table I, and incorrect answer choices selected by 20% or more students are indicated by the red font. Table I also shows the average CSEM-related PCK scores of the TAs.

### B. Results relevant to the research goal

In this section, we group questions together based on the concepts involved. We discuss questions in which few TAs identified a MCI as well as questions in which the TAs performed quite well in identifying the MCI.

#### 1. Charge distribution on conductors and insulators (Q1, Q2)

Q1 and Q2 ask about what happens to an excess charge placed at some point $P$ on a conducting (Q1) or insulating hollow sphere (Q2). For Q1, students' MCI was that the charge distributes everywhere on the inside and outside of the metal sphere (choice C, roughly one-fourth of the students). This implies that students may be thinking that the positive charges spread as far from each other as possible [23]. More than half of the TAs identified this choice showing that they are aware that students may struggle with the fact that charges on a metallic sphere are distributed only on the outer surface in equilibrium. On Q2, students had two alternate conceptions: that the charge distributes itself everywhere on the outside of the sphere, i.e., not distinguishing between insulating and conducting (choice B) and that there will be no charge left (choice E)— roughly one-fifth of the students selected each. On both Q1 and Q2, many TAs expected that the MCI is choice D for both questions, namely, that most of the charge is at point $P$, but some of it will spread over the sphere. On Q1, some of the TAs reasoned that choice D would be the MCI because students would expect that the charges would move, but that there is not enough force to move all the charges everywhere around the sphere, or that it takes more than a few seconds for the charge to spread everywhere and therefore some will remain at point $P$. For example, one interviewed TA stated: "They [students] don't expect that for a metal [there is] enough push in order to move all the charges from that point [$P$]." Another interviewed TA motivated his selection of choice D as the MCI by stating: "Most people probably think it's D […] because they might not recognize that it has to be an instantaneous distribution of charge. So, they recognize that the charge will have to spread over the surface, and since we know it's metal, I'm assuming they understand a conductor won't have charge on the inside. It [charge] is all gonna be on the surface, but they might assume that the majority of the charge hasn't fully distributed yet." On Q2, TAs' most common reasoning for selecting choice D was that it was the incorrect answer choice that is most similar to the correct one and that students may have some understanding that an insulating sphere is different from a conducting sphere, but would not fully understand it. For example, one TA said: "If they understand this is insulating material [i.e., they do not miss this information when reading the question], they will choose D […] because they know something about insulating that it is not like the conducting, but they [may not know] that the charge will stay at the position [$P$]." This seems reasonable; however, it appears that few students selected this answer choice.

#### 2. Coulomb's force law (Q3, Q4, Q5, Q6, Q7, Q8)

Q3, Q4, and Q5 are related and shown in Fig. 1. On Q3, three-fourths of the students realize that the force on the $+Q$ charge should increase by a factor of 4. On Q3 and Q6 the largest percentage of students who selected an incorrect answer choice was 13%, so it does not seem like there are persistent alternate conceptions on these questions.

For questions 3 -5:
Two small objects each with a net charge of +Q exert a force of magnitude F on each other.

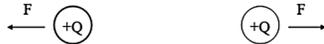

We replace one of the objects with another whose net charge is +4Q:

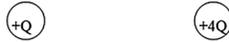

3. The original magnitude of the force on the +Q charge was F; what is the magnitude of the force on the +Q now?

   (a) 16F      (b) 4F      (c) F      (d) F/4      (e) other

4. What is the magnitude of the force on the +4Q charge?

   (a) 16F      (b) 4F      (c) F      (d) F/4      (e) other

Next we move the +Q and +4Q charges to be 3 times as far apart as they were:

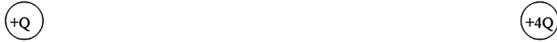

5. Now what is the magnitude of the force on the +4Q?

   (a) F/9      (b) F/3      (c) 4F/9      (d) 4F/3      (e) other

FIG. 1. Questions 3, 4, and 5 on the CSEM.

On Q4, about one-third of the students think that after increasing the charge on the right from $+Q$ to $+4Q$, the magnitude of the force on it remains the same, $F$ (instead of increasing by a factor of 4 to $4F$). This suggests that they might have the alternate conception that the electric force on a charge is only proportional to the charge that is applying the force [23]. Students may also not recognize that Newton's 3rd law applies (i.e., the electric force exerted on the $+Q$ charge by the $+4Q$ charge has the same magnitude as the electric force exerted on the $+4Q$ charge by the $+Q$ charge). This alternate conception was selected by 57% of the TAs.

Q5 asks students what happens to the magnitude of the force when the charges are moved to be 3 times as far apart. One fifth of the students who selected choice C on Q4 thought that the force will now decrease by a factor of 3 and selected choice B on Q5, while a smaller percentage (14%) thought that the force will decrease by a factor of 9 (correct thinking, but incorrect conclusion because the force on the $+4Q$ charge is initially $4F$ not $F$). In other words, the MCI is that when the two charges are moved three times as far apart, the force on them decreases by a factor of 3. If the TAs are aware that this is the MCI, then among the TAs who selected choice C on Q4, many of them should select choice B on Q5. However, while 57% of the TAs selected choice C on Q4, only one-sixth of the TAs identified choice B as the MCI on Q5, and one-third selected choice A, possibly because choice A is a combination of a correct idea (force decreases by a factor of 9) and an incorrect one (force on +4Q charge before increasing the distance between the two charges is $F$).

On Q5, roughly one-fourth to one-fifth of the students selected option D and option B, each. The students who selected either of these two options are likely to think that the electric force is inversely proportional to the distance (instead of distance squared), so that when the separation between two charges is tripled, the force between them decreases by a factor of 3. So, if a student answers $4F/3$ on Q5, they probably thought that the force decreased by a factor of 3, and the original force was $4F$ (Q4). If instead, a student answers $F/3$ on Q5, that student probably thought that the original force was $F$. Half of the TAs identified option D as the MCI, whereas less than one-sixth of the students selected option B. On Q4 and Q5, the TAs identified the MCIs quite well.

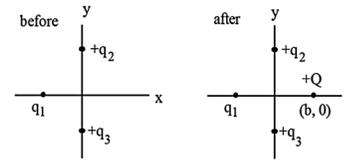

FIG. 2. Figure provided for Q8 on the CSEM.

On Q7, there is one strong distractor (choice C, more than 40% of students); another choice (A) is selected by one-fifth of the students. Nearly identical percentages of TAs (between 40% and 45%) selected choices C and A, respectively, as the MCI. However, choice C is much more common than choice A among students. Q8 provides students with the two situations depicted in Fig. 2 and states that in the configuration on the left, charges $q_2$ and $q_3$ are positive and that the net force acting on $q_1$ as the result of its interaction with the two charges points in the positive $x$ direction (to the right). The question asks what happens to the force acting on $q_1$ when another positive charge $(+Q)$ is placed at the location shown in the configuration on the right. The MCI (choice D) selected by roughly one-fifth of students is that the force will increase and its direction may change due to the interaction between $Q$ and charges $q_2$ and $q_3$. While 43% selected this as the MCI, more than one-fourth of the TAs selected choice E, which states that the answer cannot be determined without knowing the magnitude of $q_1$ and/or $Q$. However, this answer choice was selected by less than 10% of students. In interviews, some of the TAs also selected choice E as the MCI. One interviewed TA, for example, motivated selecting choice E by stating: "I think most of them [students] will go with E […] because they might think that $F$ is $kq_1q_2$ divided by $r$ [squared] and then they think, 'ok, nothing is [given], $q$ is not [given], $r$ is not [given]', then they cannot decide [what happens to] the force." It appears that some of the TAs think that students may remember the equation for the electric force acting between two charges, but since none of the information is explicitly given (i.e., by providing values for the charges and distances), the electric force cannot be calculated so the question cannot be answered. However, it appears that very few students may be reasoning this way since less than 10% selected this answer choice.

### 3. Connection between electric field and electric force (Q10, Q12, Q15)

Q10 states that a positive charge is released from rest in a uniform electric field and asks about its subsequent motion. The two MCIs are that the charge remains at rest (choice E, selected by one-fourth of the students and less than 10% of TAs) and that it will move at constant velocity (choice B, selected by one-fifth of the students). In the interviews, TAs were explicitly asked whether they expected that students would harbor the alternate conception of choosing choice E. Nearly all the interviewed TAs said that it is unlikely that students do not know that charges placed in an electric field would move and, thus, the interviews highlighted how challenging it is for TAs to identify this alternate conception.

Answer choice A is similar to choice B except that it says that the charge moves at a constant speed instead of constant velocity and only 6% of the students selected this answer choice. However, one-fourth of TAs selected it. During the interviews, some of the TAs who selected this answer choice did not seem to consider B very carefully. For example, one TA stated: "They might think it will go at constant speed because the field is uniform so the effect is constant throughout the path." However, a more common occurrence in the interviews was for TAs to consider both choices A and B as the MCIs and either say they are not sure which one is more common or that students would select among these two answer choices equally. It is possible that in the quantitative study conducted in the TA professional development course, TAs had similar considerations and some TAs opted for choice A while others opted for choice B as the MCI. However, as shown in Table I, much fewer students selected choice A compared to choice B. There are two other questions in this grouping: Q12, which does not appear to have any common alternate conceptions (largest percentage of students who selected an incorrect answer choice is 13%) and Q15 (shown in Fig. 3) on which the TAs performed well at identifying them (91% of the TAs selected one of the two alternate conceptions on this question). However, on Q15, one-fourth of students expected that the electric force points directly towards the positive charge from which all the field lines originate (choice B), but only one-sixth of the TAs identified this as the MCI. This is likely due to another alternate conception common amongst more students (roughly one-third), namely, that the electric force points to the right neglecting to incorporate the sign of the charge. The vast majority of the TAs (roughly three-fourths) identified this more common alternate conception.

### 4. Induced charge and electric field or force (Q13, Q14)

Q13 and Q14 provide students with the diagrams shown in Fig. 4. In Q13, the sphere is hollow and conducting and has an excess positive charge on its surface. The question asks for the direction of the electric field at the center of the sphere. In Q14, the sphere is also hollow and conducting, but it has no excess charge, and the question asks about the forces acting on the two charges.

On both of these questions, the students' MCI is to not recognize that the conducting sphere alters the electric field or forces. Thus, on Q13, roughly one-fourth of them selected choice A on which the electric field is to the left which does not incorporate the effect of the metal sphere on the electric field (as though the sphere is not present). Roughly half of the TAs selected option A as the MCI, thus suggesting that they are aware that students have difficulty recognizing how conducting objects respond to the external electric field (i.e., free charge moves in order to make the electric field inside the conductor zero).

On Q14, roughly half of the TAs selected choice A which says that the forces the two charges feel are the same (once again, as though the sphere does not affect the forces). On this question, roughly half the TAs selected other answer choices (B, C, and E), which combined were selected by only one-fourth of students. In the interviews, TAs who selected choice A as the MCI on Q13 usually did so because they expected that students would only think about the electric field caused by the $+Q$ charge and ignore the metal sphere. For example, one TA who selected A said: "Maybe someone would say leftward because they think of the positive being the source so they think of it making a [field] line and the [field] line is going outward from the charge, and they think it's just going to go straight through the sphere." On the other hand, on Q14, this same TA said that students' MCI would be choice E because they may think that the charge distribution on the sphere affects the forces that the two charges are experiencing. "They might think that little $q$ at the center of the sphere […] is feeling forces from the charges that are distributed along the surface [of the sphere], and big $Q$ here might feel force from this guy [$q$] and all the surface charges [on the sphere]." Other interviewed TAs cited similar reasoning for selecting choice E.

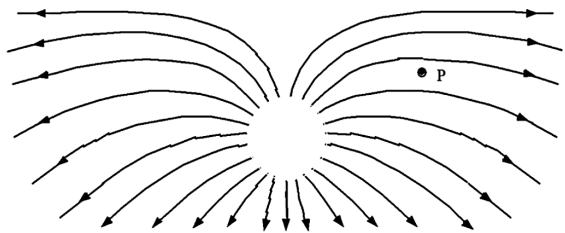

FIG. 3. Q15 on the CSEM.

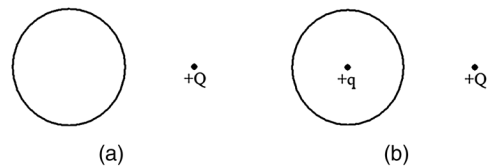

FIG. 4. Diagrams provided for Q13 (a) and Q14 (b) on the CSEM.

On Q13, another TA selected choice A as the MCI and stated that students may ignore the effect of the sphere. When looking at Q14, this TA explicitly mentioned his previous answer and stated: "My thought is similar to the last one, to kind of just ignore the sphere. So, A, maybe." In other words, students may ignore the effect of the sphere and select A. But after noticing choice E, he changed his mind and went on to say the following: "I think E [may be the MCI] because they might realize that the sphere does do something to change things, so they think 'ok, I know [the forces would normally be] equal and opposite, but now there's a sphere here, so I don't know exactly how that works' [i.e., what the effect of the sphere is] so they'll just throw in something [i.e., include some effect due to the sphere], so E is that something." It appears that this TA was aware that students may be guided by similar incorrect thinking (conducting sphere will not have an effect) on Q14 as on Q13, but on Q14, selected the answer choice which incorporates a correct idea (conducting sphere has an effect), but is missing another idea in order to be fully correct. In many other questions, TAs often selected answer choices which fit this category. For example, as mentioned earlier, on Q1, some TAs thought that students would select choice D, which states that some of the charge does spread over the sphere—a partially correct answer. Similarly, on Q2, some TAs selected the same answer, which is partially correct because some of the charge does remain at point $P$. They also sometimes explicitly noted that they were selecting this answer choice as the MCI because it is the one that is most similar to the correct answer. On Q10, many interviewed TAs considered answer choices A and B, stating they expected that students would be aware that the charge should move, but they may not know that it moves with a constant acceleration (more examples will be discussed below). While sometimes using this strategy to identify the MCI may provide a reasonable answer choice (i.e., one that is fairly common among students), it often misled the TAs into selecting an answer choice that was not very common—as was the case on Q1, Q2, and Q14 (and other questions that are discussed below). On Q14, for example, this reasoning led some TAs to select choice E as the MCI. However, this answer choice was only selected by less than one-seventh of students.

On Q14, one interviewed TA who selected choice B as the MCI to Q14 noted that students may reason in the following way: "Inside the conductor there is no field. But they might think the sphere is shielding the field due to the inside charge also. So, everything is shielded and there is no force [i.e., neither $+q$ nor $+Q$ experience a force]." Other TAs who selected choice B used very similar reasoning.

Similar to TAs' reasoning for selecting choice E discussed earlier, choice B also incorporates a partially correct idea: the metal sphere "protects" the charge inside from the effect of outside charges, which is partly why many interviewed TAs selected it as the MCI.

### 5. Connection between electric potential and electric field or force (Q16, Q18, Q19, Q20)

Q16 states that an electron is placed at a position on the $x$ axis where the electric potential is equal to $+10$ V and asks about the subsequent motion of the electron. The students' responses are spread over the four incorrect choices almost evenly. Roughly one-fourth of students thought that the electron would move towards the right (the MCI), but this answer choice was the one least likely to be selected by the TAs (less than one-sixth selected it). One interviewed TA thought that the students will place the electron on the positive $x$ axis and a positive charge at the origin of the coordinate axis (to give concreteness to the situation) and claim that the electron would move to the left.

Q18 relates to the three situations shown in Fig. 5 and asks students to compare the magnitude of the electric field at point $B$ in all three cases. Here, more than one-fourth of students selected E which states that the electric fields are equal. These students only considered that the equipotential line on which $B$ lies is at 40 V and did not recognize that it is the *change* in electric potential (i.e., gradient) that is related to the magnitude of the electric field rather than the potential itself. Just over half the TAs identified this difficulty.

Q19 asks students for the direction of the electric force acting on a positive charge if placed at point $A$ or $B$ in situation III. One-quarter of students selected choice B (right at point $A$ and right at point $B$), possibly because "right" is the direction in which the electric potential increases and they expected that a positive charge would be pushed in that direction [23]. This alternate conception was identified by 61% of the TAs. On Q20, the TAs appear to be able to identify the alternate conceptions.

### 6. Work or electric potential energy (Q11, Q17)

Q11 asks what happens to the electric potential energy of a positive charge after being released from rest in a uniform electric field. Students' MCI is that it remains constant because the electric field is uniform (choice A selected by 30% of students). A much less common answer choice is choice C; namely, that the electric potential energy will increase because the charge will move in the direction of the electric field (selected by only 13% of students). However, more than one-fourth of the TAs selected this answer choice, and during an interview, one TA reasoned that perhaps students are thinking about total energy instead of electric potential energy (or perhaps they are confusing kinetic and electric potential energy): "It (the charge) has an acceleration and velocity is increasing right? So, they [students] may think that the potential [energy] should increase because velocity is increasing." Another interviewed TA who selected choice C as the MCI selected it for a very similar reason. On Q17, students are asked to compare the work needed to move a positive charge from point $A$ to point $B$ in three different situations (shown in Fig. 5). More than one-fifth of the students answered that the most work is done when moving the charge in situation III. These students likely thought that the work is maximum in situation III because the distance over which the charge is moved is largest and did not consider the potential difference between the two points. Many TAs (63%) identified this alternate conception. On Q17, 81% of the TAs selected one of the two MCIs, thus they performed well at identifying the alternate conceptions on this question.

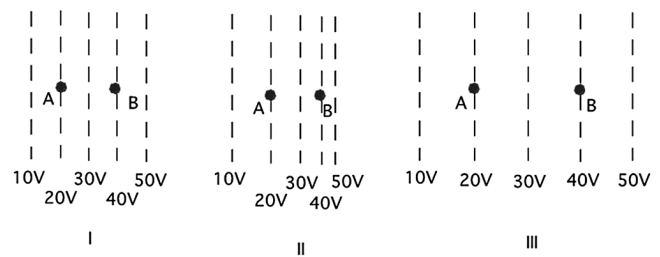

FIG. 5. Diagram provided for Q17–Q19 on the CSEM.

### 7. Force on or motion of charged particle in a magnetic field (Q21, Q22, Q25, Q27)

Q21 asks what happens to a positive charge that is placed at rest in a magnetic field. Students' MCI is that the charge moves in a circle at constant speed (choice C selected by roughly one-fifth of students). On this question, many TAs thought that students may confuse electric and magnetic field and thereby conclude that the charge moves with constant acceleration (choice B selected by 31% of TAs), but this answer choice is very rarely selected by students (less than 10% of them selected this choice). For example, one interviewed TA stated, "I can see people confusing or essentially just ignoring that it's a magnetic field thinking that it should do the same thing as it does in an electric field, so, constant acceleration. Yea, that would be my guess—B, they would think that it would do the same thing it does in an electric field."

Q22 provides the diagram shown in Fig. 6 and asks for the direction of the magnetic field responsible for making the electron path curve in the way shown. More than one-fourth of the students selected "into the page" which would be correct if the electron was positively charged and 59% of the TAs identified this difficulty. Also, 22% of the students selected upward, suggesting that they may think that the direction of the magnetic force is the same as the direction of the magnetic field, less than one-fourth of the TAs identified this alternate conception.

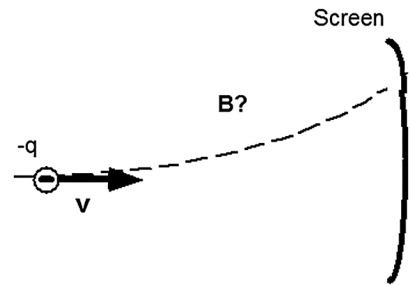

FIG. 6. Diagram provided for Q22 on the CSEM.

Q25 provides the three situations shown in Fig. 7 of a positive charge moving in an external magnetic field and asks students to rank them according to the magnitude of the magnetic force. TAs' selections are quite varied, with a significant percentage of them opting for each incorrect answer choice, suggesting they had difficulty identifying the MCIs.

Interviews suggest that TAs struggled to identify the MCI, which is that the force is largest in situation II (where the charge moves "against" the magnetic field) and least in situation III (where the charge moves "with" the electric field), and situation II is in between—choice C selected by one-fifth of students. TAs had difficulty determining how students may reason about this question incorrectly. The TAs sometimes opted for choice A (same force in all situations) because they expected that some students may only recall $qvB$ as the magnetic force on a charge moving in a magnetic field and thus conclude that the forces are equal in the three situations. If they did not select this answer choice, they usually started by stating that when the velocity and magnetic field are in the same direction, students may think that this leads to the largest force. For example, one TA stated: "They [students] are thinking "oh, the magnetic field is pushing it along in this direction and it's already moving in that direction" so that's just compounding the effect (i.e., force is largest in situation III)."

Other interviewed TAs reasoned in a similar way, but after concluding that students may think the force is largest in situation III, they had difficulty applying the same reasoning to situations I and II. They sometimes stated that for situation II, students may think that the acceleration is least because the charge is moving in a direction (partly) opposite to the magnetic field and conclude that the force is least in situation II (and select B). Other TAs stated that perhaps students are somehow thinking of the dot product instead of the cross product and conclude that choice E is the MCI. Yet other TAs, after considering situation II, changed their minds because they thought that since the charge is moving against the magnetic field, students may think that the field is exerting the largest force. This was one of the questions on the CSEM which took the TAs the most time to answer (i.e., determine what they expected would be the MCI). One TA, after trying to figure it out for a while, just gave up and said that maybe students would just rank the situations in the opposite order (i.e., students would not read the question correctly and think the situations should be ranked from least to greatest). Thus, both the interviews and the quantitative data suggest that the TAs had difficulty identifying students' alternate conception on this question.

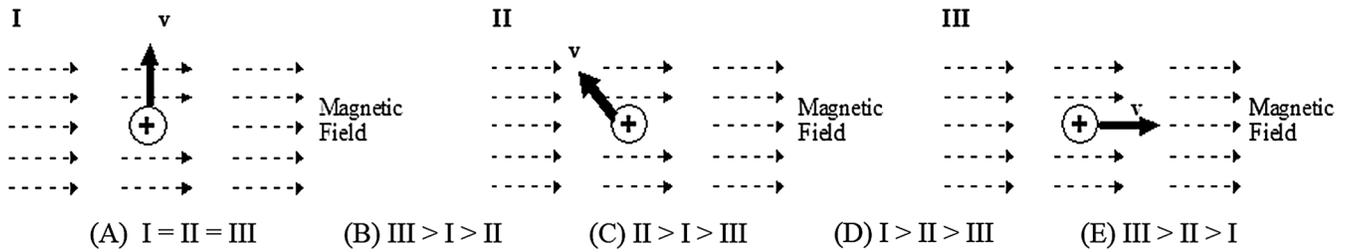

FIG. 7. Three situations and answer choices provided in Q25 on the CSEM.

Q27 shows a positive charge placed at rest near two magnets, the one on the left being 3 stronger than the one on the right (see Fig. 8). It asks for the magnetic force acting on the charge and provides the answer choices shown in Fig. 8. On this question, the MCIs are choice A (∼one-fifth) and choice D (∼one-fourth). While 46% of the TAs selected choice A, only 12% of the TAs selected choice D, and one-fourth selected choice C which is selected by less than 10% of students. One interviewed TA selected choice C because he expected students to think that the magnet on the left is pushing the charge towards the right and the magnet on the right is pushing the charge towards the left. When asked why he expected students to think this way he stated that he did not know how to explain it, it was just his gut feeling based on his experience teaching recitations.

### 8. Magnetic field caused by a current (Q23, Q26, Q28)

On Q26 (shown in Fig. 9), students' MCI is that the magnetic field is radially outward from the wire (choice D, one-fifth of the students). On this question, roughly half of the TAs selected choice C in which the direction of the magnetic field is opposite to the correct direction (i.e., clockwise instead of counterclockwise), but only 6% of students selected this answer choice. All the interviewed TAs who selected this answer choice essentially said that students may either use their left hand or use the right-hand rule incorrectly. However, the choices selected by students do not suggest this as a major difficulty.

On Q26, some interviewed TAs used similar reasoning as some of the TAs who selected choice E on Q14— students have some correct ideas (try to use the right-hand rule), but are not fully correct (obtain the incorrect direction). It is important to point out that after recognizing that students may be answering the question incorrectly for this reason (which does not seem to be common), the interviewed TAs did not consider all the other answer choices carefully and did not realize that students may have other alternate conceptions, namely, that the magnetic field would be radially outward from the wire (i.e., confusion between electric and magnetic field). After the TAs answered all the other questions in the interview, they were often asked to return to this question and think about whether they expected that any students would select choice D (radially outward magnetic field). After being asked to consider this answer choice explicitly, they were often able to recognize the alternate conception guiding students to select choice D and some interviewed TAs wanted to change their original answer. Similarly to Q14, some TAs attempted to identify common alternate conceptions on Q26 by arguing that students may have some correct ideas, but miss something that causes them to not have the fully correct answer. However, it appears that for this question (and others mentioned earlier), this type of reasoning from the TAs often steered them in the wrong direction and caused them to identify an answer choice that is not common among students while missing the MCI.

On Q28 (shown in Fig. 10), the loops carry currents of equal magnitude and the question asks for the direction of the magnetic field at point $P$. The MCI is that the two magnetic fields created by the two wires cancel out and the magnetic field at point $P$ is zero (choice E), selected by roughly one-third of students. These students likely thought that the magnetic fields created by the two loops are in opposite directions and they therefore cancel [23]. This alternate conception was identified by 55% of the TAs, but one-third of them also selected choice A (selected by less than 10% of students). Similarly to Q26 discussed above, all of the TAs who selected this answer choice during interviews claimed that students may use the right-hand rule incorrectly and obtain the incorrect direction, however, it appears that very few students do this.

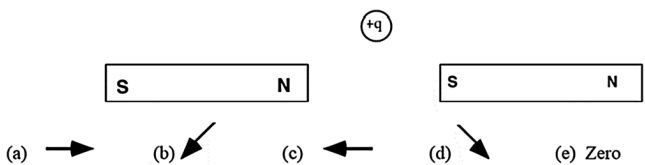

FIG. 8. Physical situation and answers provided for Q27 on the CSEM.

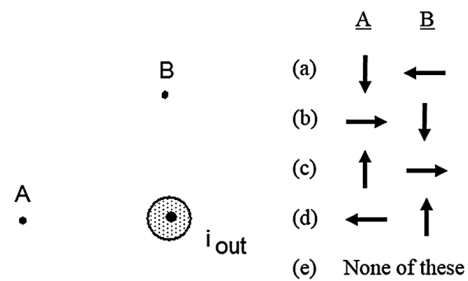

FIG. 9. Diagram and answer choices for Q26 on the CSEM.

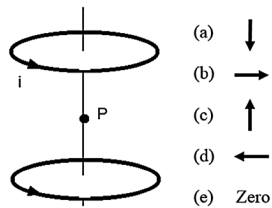

FIG. 10. Diagram and answer choices for Q28 on the CSEM.

### 9. Faraday's law (Q29, Q30, Q31, Q32)

Q29 asks students to identify all of the situations shown in Fig. 11 in which the light bulb is glowing. Roughly one-fourth of students only selected situations I and IV in which there is relative motion between the magnet and the loop. These students did not recognize that in situation II, the electric flux is changing (because the area of the loop is changing) and therefore there will be an induced emf (electromotive force) in the loop (light bulb glows). Roughly half the TAs identified this alternate conception. Furthermore, roughly one-fourth of the students also selected situation III (i.e., answered that the light bulb glows in situations I, III and IV, choice A), sometimes due to overgeneralizing that there is an induced emf in any

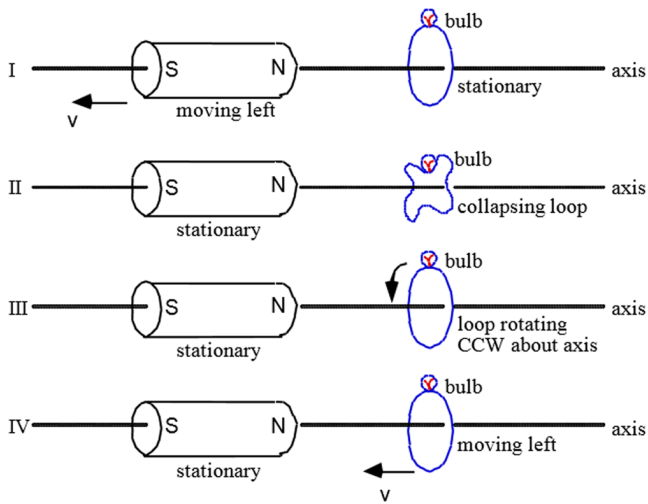

FIG. 11. Diagram provided for Q29 on the CSEM.

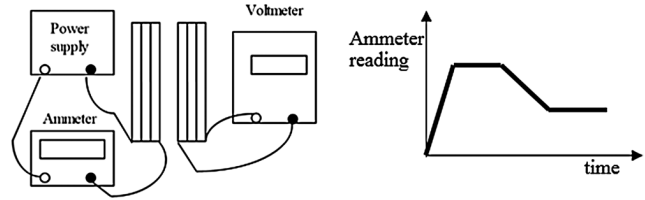

Which of the following graphs correctly shows the time dependence of the voltmeter reading?

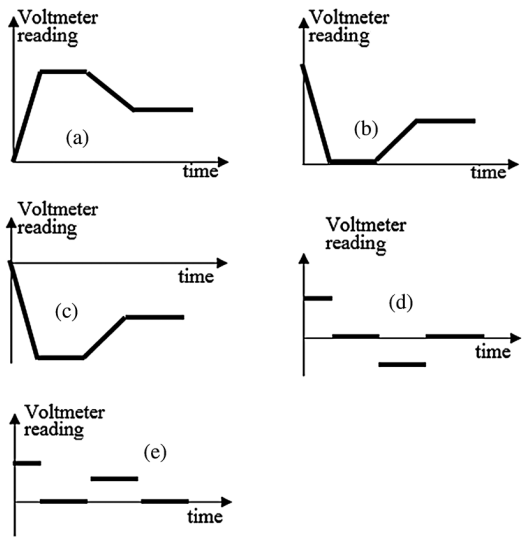

FIG. 12. Q32 on the CSEM.

situation in which the loop is moving, while one-fourth of the TAs identified this alternate conception.

On Q29 and Q30, 77% and 81% of the TAs identified one of the two MCIs in each question, while on Q31, students seem to be randomly selecting from the four incorrect answer choices.

Q32 is one of the most challenging questions on the CSEM (less than one-fifth of the students answered it correctly). The question and answers are shown in Fig. 12. On this question, the MCI is choice B (selected by 40% of students). The corresponding alternate conception is that the reading on the voltmeter opposes the reading in the ammeter (i.e., reading on the ammeter increases, therefore reading on the voltmeter decreases and vice versa). The students may be trying to apply Lenz's law, but may not realize that the induced emf opposes the *change* in flux rather than the flux itself. It indicates that they have a lot of difficulty recognizing that the induced emf in the secondary coil is only nonzero when the current in the primary coil is changing. However, it appears that many TAs are unaware of this difficulty. Roughly 10% of TAs identified this alternate conception. On the other hand, 31% of the TAs selected choice E, but only 1% of students selected this choice. In the interviews, one TA selected this choice, explaining that it is possible that students only think of the magnitude of the emf (once again, the TA combined a correct idea, i.e., only a changing flux induces an emf, with an incorrect one, i.e., students do not recognize that the induced emf changes direction).

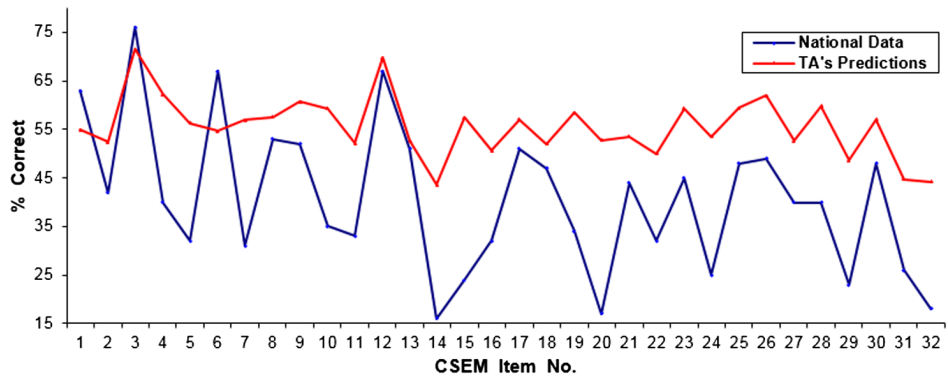

FIG. 13. Comparison of the percentages of correct answers predicted by TAs with algebra-based students' actual performance after traditional instruction as obtained from Ref. [23]. Standard deviations range between 17.7 and 24.6 and are not shown for clarity.

### 10. To what extent are TAs able to predict the difficulty of the questions?

Figure 13 shows TAs' average predictions of the difficulty of each question on the CSEM, i.e., the percentage of students who answered each question correctly (TAs' predictions) as well as the actual difficulty of each question (National Data in Ref. [23]). Figure 13 shows that the TAs underestimated the average difficulty of the majority of the questions on the CSEM. The discrepancy between TAs' predicted difficulty and the actual difficulty is quite large for some questions, in particular, the questions that were most difficult for students (e.g., Q14, Q20, Q24, Q29, Q31, Q32). Figure 13 also shows that TAs' predicted difficulty does not fluctuate very much: with the exception of only five questions, the TAs' predicted difficulty is between 45% and 65% for all the questions on the CSEM, thus indicating that the TAs did not have a good sense of how difficult the questions are from the perspective of students. This conclusion is further supported by averaging TAs' predictions over all questions and comparing them to the actual average difficulty: TAs overpredicted students' performance on the CSEM by 15% on average.

## IV. USING A PCK TASK AS A PEDAGOGICAL TOOL

Many TAs explicitly noted that the CSEM-related PCK task was challenging and it was difficult for them to think about the difficulty of the physics questions from a student's perspective. In the think-aloud interviews, graduate students sometimes made comments which indicated that they found the task challenging (e.g., explicitly commenting "I don't know students well enough…"). However, many TAs noted that the CSEM-related PCK task was worthwhile and helped them think about the importance of putting themselves in their students' shoes in order for teaching and learning to be effective, especially after receiving student data on how students actually performed and discussing particular student alternate conceptions.

Our interviews suggest that if such a task is used for TA professional development, it is best for teaching assistants to be explicitly told to first try to identify (and perhaps write down) what alternate conceptions or incorrect reasoning may lead students to select each of the incorrect answer choices before deciding which one is the MCI. In think-aloud interviews, we found that when TAs were explicitly prompted to consider all alternative answer choices and articulate why a student may select each, they were very reflective and often managed to identify the MCI. If there is not sufficient time for this process during the professional development activity, either the TAs can be asked to perform this task as homework before the discussion during the professional development activity or the professional development leaders can select a subset of questions that would be most productive for discussion based upon the study described here. For example, our data suggest that Q2 would be a good question to discuss. First, the TAs should be asked to identify the most common incorrect answer choice (after they are either provided with the correct answer or they are asked to identify it), and our quantitative data suggests that most TAs will select D as the MCI (only selected by 11% of students). However, other TAs will select B which is a common alternate conception. After working on this task, the TAs could be asked to convince one another that their choice is actually the MCI of students, and the professional development leader can guide the discussion. Finally, the TAs could be shown the student data and asked to reflect upon it. There are many other questions on the CSEM in which a significant fraction of TAs selected an answer that is not common, while other TAs selected the MCI, e.g., questions Q4, Q7, Q8, Q11, Q14, Q17, Q21, Q24, Q25, Q26, Q27, Q28, Q32. These types of questions can be used in a similar manner to what is discussed above.

Other questions could be used to build TA confidence. These are the questions in which the majority of TAs selected the MCI. For example, Q17 in which 63% of the TAs selected the MCI. Questions which can be used in this manner are Q4, Q5, Q7, Q13, Q15, Q17, Q20, Q22, Q24, Q29.

We also note that sometimes the same question can be used in two different ways. For example, for Q4, our quantitative data suggest that the majority of the TAs select the two MCIs (C and D). After they do this in a professional development class, the instructor can offer praise that they had identified the two most common student alternate conceptions and follow through with "Now, let's think about which one is more common. Do you expect that C is more common, or D, or perhaps similar percentages of students select either choice? Discuss with each other and predict the percentage of students who select C or D in the post-test."

Furthermore, our qualitative data provide reasons the TAs sometimes use when selecting certain answer choices as the MCIs when those answer choices are not actually common among students. For example, on Q26, TAs often select the answer choice in which the direction of the electric field is opposite to the correct one, motivating their choice by saying that perhaps students will use the right-hand rule incorrectly, or use their left hand. In a professional development class, after the TAs mention this reasoning, the professional development leader can ask them to think of any other incorrect reasoning students may use. If the TAs struggle, they can be directed to think about answer choice D, and our interviews suggest that the TAs will likely be able to figure out that students may think the magnetic field radiates outward (similar to electric field). After this they can be asked what they expect would be the most common incorrect reasoning used by students and again, our qualitative interviews suggest that the TAs will likely identify the latter incorrect reasoning (field radiates outward) as more common than the former (using the left hand, or using the right-hand rule incorrectly).

In addition, our research suggests that the TAs are usually thoughtful when thinking aloud about this PCK task, thus, it will be useful for the TAs to reflect upon this task with their peers during the TA development activity even if they did not manage to identify the MCIs of students very well.

We note that two of the authors (A. M. and C. S.) have been using tasks similar to the one described here in the professional development of TAs at their institutions and have found them to be very useful in setting the stage for a discussion on the importance of being aware of students' difficulties and alternate conceptions in order to design instruction to help students learn. The TAs discuss questions which have been carefully selected to engender productive discussions among TAs as discussed above. The TAs are explicitly asked to identify and discuss with each other what reasoning students may use to select each incorrect answer choice before making a decision about which one is the MCI. Additionally, since only a subset of questions is selected, there is more time for the TAs to also spend predicting the difficulty of each question. After the TAs complete the task, they are shown data from students, and some TAs explicitly express that it is very valuable for them to learn about the common student difficulties in concrete contexts. We found that TAs tend to trust student data more than statements like "research has found that…" The discussion is then focused on how TAs can identify common student difficulties related to various physics concepts, e.g., by listening to students when reasoning about physics and coming up with guiding questions in real time to develop a grasp of how students are thinking in specific contexts. At one of the institutions (University of Cinncinnati), the rest of the professional development program (which meets once a week for a semester) is focused on the tutorials students work on and their common difficulties on specific questions on the tutorials, as well as effective approaches the TAs can use to help students develop a coherent knowledge structure of those physics concepts. Using such tasks with actual data from students for TA professional development can be effective at other institutions as well.

## V. DISCUSSION AND SUMMARY

Awareness of students' common difficulties and being able to understand how challenging certain concepts are for students are important aspects of pedagogical content knowledge. One can take advantage of knowledge of students' common difficulties and use them as resources to design effective pedagogical approaches to help students learn better [44,46,50]. Our investigation used the CSEM to evaluate this aspect of pedagogical content knowledge in the context of electricity and magnetism for 81 TAs who were all first-year physics graduate students enrolled in a TA professional development course. For each item on the CSEM, the TAs were asked to identify what they expect is the MCI of introductory students after traditional instruction. Additionally, in years two and three of the study, the TAs were also asked to estimate the difficulty of each question on the CSEM. In all three years there was an in-class discussion with the TAs related to the PCK task. Additionally, think-aloud interviews were conducted to obtain an in-depth account of what reasoning TAs use to arrive at the conclusion that certain alternate conceptions may be common.

### 1. General approach often used by the TAs to identify common incorrect answer choices of students

When trying to decide what answer choices would be common among students, TAs often selected answer choices which incorporate both correct and incorrect ideas. While this approach was sometimes productive in helping them identify the MCIs, it often led to TAs selecting answer choices that were not common at all. Also, after TAs identified a particular answer choice which incorporated a correct and incorrect idea, they sometimes neglected to consider other answer choices carefully or think about what alternate conceptions could lead to students selecting them. There are many examples, for instance, Q26 (shown in Fig. 9) in which TAs often selected the answer choice which has the direction opposite to the correct direction. They often stated that they were motivated to select this choice because students may try to use the right-hand rule (correct idea), but do so incorrectly. We note that after they were explicitly asked to think about a particular answer choice (e.g., choice D for Q26), the TAs sometimes predicted the alternate conception, in this case that the magnetic field points radially outward from the wire (i.e., making a generalization from electric field due to a positive charge, e.g., point charge or line of charge), and stated that they expected this answer choice to be more common than the one they originally selected.

### 2. TAs struggled to identify alternate conceptions regarding how charge distributes on conductors and insulators

There are two questions on the CSEM that ask what happens to a charge placed at a particular point on a conducting or insulating sphere. For both questions, many TAs selected answer choices that were not common among students. On the question in which the sphere is insulating, nearly half the TAs expected that students would think that most of the charge remains where it was placed, but some does spread over the sphere. Interviews suggested that the TAs selected this answer choice because it is the choice which is most similar to the correct answer (charge remains where it was placed), i.e., the TAs used the same strategy we described above in other contexts.

### 3. TAs struggled to identify alternate conceptions regarding the magnetic field caused by a current

On both questions related to magnetic field caused by a current for which there was a common alternate conception, the TAs selected answer choices which are not at all common among students. On both questions TAs' often selected answer choices in which the right-hand rule was used incorrectly, but very few students selected those answer choices.

### 4. TAs struggled to identify alternate conceptions regarding the motion of or force on a charged particle in a magnetic field

Out of the four questions dealing with the concept of Lorenz force (Q21, Q22, Q25, Q27), only on one of them (Q22) did the majority of TAs identify the MCIs. On the other ones, the TAs often selected answer choices that were not common. Also, Q25 was one of the most challenging questions for the TAs; in interviews they often spent a considerable amount of time trying to figure out how students may answer the question and sometimes even ended up essentially guessing, or committing to an answer only after being asked to select one.

### 5. Alternate conceptions held by very few students which TAs expected would be the MCIs

There were multiple instances in which TAs selected certain incorrect answer choices which they thought would be MCIs among students, but those answer choices were very rarely selected by students. Three such examples are presented in the preceding paragraphs and there are many others. For example, on Q21, 31% of the TAs expected that students would confuse the magnetic field with an electric field and think that the charge will move at constant acceleration, but only 8% of students selected this answer choice and on Q32, 31% of the TAs selected choice E, but only 1% of the students selected this option.

### 6. Alternate conceptions that the TAs were able to identify

The TAs performed reasonably well at identifying alternate conceptions related to Coulomb's force law (Q3–Q8), although there is room for improvement, especially on Q8. On Q3 and Q6, there are no strong alternate conceptions, and on Q4, Q5, and Q7 the majority of the TAs identified the common alternate conceptions. Q8 is the only one on which the TAs could improve significantly, and this is the only question of the group that has a complicated setup and asks students to compare two configurations side by side, one with three charges and the other with four. It is possible that TAs' lower performance in identifying the MCIs on this question was due to the setup being more complicated than those used in the other questions. TAs performed reasonably well in identifying the alternate conception that the electric field inside a hollow metallic sphere due to an external charge is the same as it would be without the hollow metal sphere. In other words, TAs were aware that students have difficulty understanding that the inside of a metallic sphere is shielded from outside electric fields. On two other questions involving Faraday's law and Lenz's law (Q29 and Q30), TAs performed well in identifying students' alternate conceptions.

### 7. TAs' ability to predict the difficulty of the questions on the CSEM

Our results also suggest that the TAs typically underestimated the difficulty of the questions on the CSEM, especially on the challenging questions. For all but five questions on the CSEM, TAs' average predictions for the percentage of students who answer the questions correctly were between 45% and 65%, while the actual percentages varied much more widely. This strongly suggests that the TAs struggled to think about the difficulty of the questions from a student's perspective.

### 8. Using a PCK task as a pedagogical tool

We have been using a PCK task as a pedagogical tool in our semester-long professional development programs, and the data collected in this study (as well as our earlier studies of PCK), has helped design effective discussions about the importance of being knowledgeable of student difficulties. Certain questions are selected for different reasons. For example, questions in which the quantitative data suggest that TAs identify the MCIs can be used to build confidence and help TAs recognize that they are knowledgeable of certain ways in which students reason. For other questions, the quantitative data suggest that TAs select two or more answer choices as most common, and these questions can lead to productive discussions as the TAs try to convince one another that a particular answer choice is more common than another, or that two answer choices are likely going to be selected by similar percentages of students. Quantitative student data should always be shared to help convince the TAs that certain incorrect answer choices are very common among students, and our experience with using a PCK task as a pedagogical tool has shown that the TAs generally appreciate learning about student difficulties in this manner.

### 9. Comparison to prior studies related to TAs' PCK for multiple choice assessments

Comparison with our earlier studies of PCK [44,46,51] using FCI and TUG-K suggests that the PCK task may be more challenging when the assessment used is the CSEM compared to other assessments related to force and motion or kinematics. One potential reason for this may be the difference between the topics of mechanics (including kinematics) and electricity and magnetism: our daily experience with the real world leads to a relatively predictable (Aristotelian) world view and TAs could more easily reason their way to common misconceptions held by students. Electricity and magnetism, on the other hand, deals with concepts that are not primarily learned experientially (e.g., charges, fields, and currents), which likely makes it more difficult to predict the MCIs of students. We note, however, that whether the context is electricity and magnetism, force and motion, kinematics, or quantum mechanics, whether intuitive or not, student difficulties can be classified in a few categories [52,53]. Knowing the types of incorrect reasoning students engage in for a particular context can be used as resources in designing instruction to help students develop a robust knowledge structure [53].

Despite the differences mentioned above, there are many commonalities in the three PCK studies. In all of these studies, there are questions on which TAs' performance at identifying common student difficulties is good, while there are also questions in which TAs struggled to identify student difficulties. Both interviews and the quantitative data show that it was often the case that TAs selected answer choices that are not very common among students. In interviews, they sometimes considered different answer choices and struggled to select the MCI, sometimes only doing so after being reminded that they should try to identify the MCI. We note that with the goal of improving TAs' PCK, in the future we plan to write up a lesson plan for using concept inventories as part of the TA professional development program and share it via a physics teaching support website [54].

Finally, our earlier studies using the TUG-K and FCI showed that the ability to identify common students' alternate conceptions was not dependent on familiarity with U.S. teaching practices and that TAs exhibited comparable performance in identifying students' alternate conceptions for the FCI or TUG-K regardless of whether they obtained their undergraduate degree in the U.S. or elsewhere. Therefore, we did not explicitly compare the PCK performance of TAs with different institutional backgrounds in this study. However, informal observations during the TA professional development course as well as interviews suggest that the CSEM related PCK performance of these TAs (e.g., from China vs US) is likely to be comparable.

### ACKNOWLEDGMENTS

We are grateful to NSF for Grant No. DUE-1524575 as well as all the TAs who participated in the studies.

---